\documentclass[12pt]{article}
\usepackage{amsmath}
\usepackage{graphicx,psfrag,epsf}
\usepackage{enumerate}
\usepackage{natbib}
\usepackage{url} 
\usepackage{amsfonts}
\usepackage{bbm}
\usepackage{color}
\usepackage{subcaption}
\usepackage[shortlabels]{enumitem}
\usepackage{booktabs}
\usepackage{rotating}

\newcommand{\blind}{0}

\DeclareMathOperator{\logit}{logit}

\newtheorem{assumption}{Assumption}

\date{} 

\begin{document}

\def\spacingset#1{\renewcommand{\baselinestretch}%
{#1}\small\normalsize} \spacingset{1}


\if0\blind
{
  \title{\bf Combining Doubly Robust Methods and Machine Learning for Estimating Average Treatment Effects for Observational Real-world Data}
  \author{
  \large{Xiaoqing Tan}\\
    \large{Department of Biostatistics, University of Pittsburgh, Pennsylvania, USA}\\
    \large{and} \\
    \large{Shu Yang}\thanks{
    Yang is partially supported by the NSF grant DMS 1811245, NIA grant 1R01AG066883, and NIEHS grant 1R01ES031651. 
    }\hspace{.2cm}\\
    \large{Department of Statistics, North Carolina State University, North Carolina, USA}\\
    \large{and} \\
    \large{Wenyu Ye,  Douglas E. Faries, Ilya Lipkovich and Zbigniew Kadziola} \\
    \large{Eli Lilly and Company, Indiana, USA}
    }
  \maketitle
} \fi

\if1\blind
{
  \bigskip
  \bigskip
  \bigskip
  \begin{center}
    {\LARGE\bf Combining Doubly Robust Methods and Machine Learning for Estimating Average Treatment Effects for Observational Real-world Data}
\end{center}
  \medskip
} \fi

\vspace{-0.5cm}
\begin{abstract}
\normalsize{Observational cohort studies are increasingly being used for comparative effectiveness research to assess the safety of therapeutics. Recently, various doubly robust methods have been proposed for average treatment effect estimation by combining the treatment model and the outcome model via different vehicles, such as matching, weighting, and regression. The key advantage of doubly robust estimators is that they require either the treatment model or the outcome model to be correctly specified to obtain a consistent estimator of average treatment effects, and therefore lead to a more accurate and often more precise inference. However, little work has been done to understand how doubly robust estimators differ due to their unique strategies of using the treatment and outcome models and how machine learning techniques can be combined to boost their performance. Here we examine multiple popular doubly robust methods and compare their performance using different treatment and outcome modeling via extensive simulations and a real-world application. We found that incorporating machine learning with doubly robust estimators such as the targeted maximum likelihood estimator gives the best overall performance. Practical guidance on how to apply doubly robust estimators is provided.}
\end{abstract}

\noindent%
\vfill

\newpage
\spacingset{1.8} 

\section{Introduction}
\label{s:intro}

Randomized control trials (RCTs) are considered to be the gold standard for establishing the causal effects of interventions. They evaluate interventions among comparable groups. However, sometime it would be impossible to conduct RCTs due to limited resources or ethical issues. Observational studies, on the other hand, examine effects in ``real world" settings without manipulation. As there is no intervention, some individuals with certain characteristics may have a different probability of being exposed to treatment than others, meaning that the covariate information between treatment groups may be highly imbalanced. Therefore, it's important to adjust for covariate imbalance issues in observational studies.

There are two ways of adjustment for observational studies. The first kind is based on the treatment model, also known as the propensity score (PS) model where the PS is defined to be the probability of being treated given covariates. The common inverse propensity treatment weighted estimator falls into this category. The idea of weighting is to create a weighted pseudo-population where treatments are ``randomized". Another kind is based on the outcome model. This outcome imputation approach tries to impute the missing potential outcomes based on outcome modeling. Estimators based on PS modeling require the correct treatment model, and estimators based on outcome modeling require the correct outcome model. 
In practice, it's common to use linear models for PS and outcome modeling. 
This, however, could be problematic because linearity may be an inappropriate assumption for PS and outcomes when the response surface is nonlinear. 
To account for potential nonlinearity, more flexible models are needed to be considered. 

Doubly robust estimators combine the above two adjustments in a fortuitous way that the causal estimator can be consistent if either the outcome model or the treatment model is correctly specified. 
Recently, various doubly robust estimators of different kinds such as weighting, matching, and regression have been proposed in the literature to estimate the average treatment effects (ATEs) \citep{glynn2010introduction,van2006targeted,yang2020multiply,zhou2019penalized}. Although they all use the PS and outcome mean models, they combine them differently. Also, doubly robust estimators especially the ones derived from the semiparametric efficiency theory are known to have the \textit{rate double robustness} property \citep{kang2007demystifying,10.1111/ectj.12097} in the sense that they remain root-$n$ consistent and asymptotically normal when using machine learning approaches to estimating the nuisance functions. The unaddressed question is how different doubly robust estimators perform coupled with machine learning approaches. Also, little work has been done to understand the challenges of covariates selection, overlapping of covariate distribution, and treatment effect heterogeneity for doubly robust estimators. 


In this paper, we review multiple popular doubly robust methods from the categories of matching, weighting, or regression, and compare their performance using different PS and outcome modeling via extensive simulations as well as a real-world application. We found that incorporating machine learning with doubly robust estimators such as the targeted maximum likelihood estimator gives the best overall performance on estimating ATEs. The main contribution of the paper is that we conduct a comprehensive evaluation of the empirical performance of estimators. Also, practical guidance in applying these estimators is provided.

The remaining paper is organized as follows. In Section \ref{s:estimator}, we discuss both doubly robust estimators and singly robust estimators in detail. 
Section \ref{s:sim} presents the extensive comparative simulations and Section \ref{s:real} reports on performance of these estimators on a real-world application. 
Section \ref{s:discuss} provides practical guidance and concludes the paper.

\section{Methodology: Singly and Doubly Robust Estimators}
\label{s:estimator}

\subsection{Notation, Assumptions, and Estimand}
\label{sec:assump}

All aforementioned methods are based on the potential outcomes framework \citep{neyman1923applications,rubin1974estimating}. Let $X_i$ be the set of observed covariates, $A_i$ be the binary treatment indicator, and $Y_i$ be the observed outcome for subject $i=1,2,\ldots,n$. Let $Y_i(a)$ be the potential outcome had subject $i$ been given treatment assignment $a$, where $a=1$ is the treatment and $a=0$ is the control. 
We assume subjects 
are independent. 
The causal estimand of interest is the average treatment effect $\tau$, which is defined as $\tau = E\{Y(1)-Y(0)\}$. 

In reality, since only one of the potential outcomes is observed and another is missing, sometimes the fundamental problem of causal inference (i.e. estimating average treatment effect) is posited as a missing data problem. 
To estimate causal parameter $\tau$ from data with non-randomized treatment assignment, the following causal assumptions are needed. 
\begin{assumption}[Stable Unit Treatment Value]\label{assump:sutva}
    The potential outcomes of any individual are unrelated to the treatment assignment of other individuals and there is no multiple versions of the treatment. 
    Sometimes this is referred to as ``consistency" in that the observed outcome is equal to the potential outcome under the actually assigned treatment, i.e., 
    the observed outcome is given by $Y = Y(A) = AY(1) + (1 - A)Y(0)$. 
\end{assumption}
\begin{assumption}[Conditional unconfoundedness or treatment ignorability]\label{assump:unconf}Given covariates $X$, treatment assignment is independent of potential outcomes, i.e. $\{Y(0), Y(1)\} \perp A | X$. 
\end{assumption}
\begin{assumption}[Overlap or common support or positivity]\label{assump:overlap} All subjects are possible to receive either arm of treatment, i.e., $0< P(A = 1| X=x) < 1   \quad \forall x \in X$.
\end{assumption}

Assumption~\ref{assump:sutva} ensures there is no interference among subjects and there is no multiple versions of the treatment. 
For observational studies, because the exposure to treatment is not controlled, treatment may be related to the way a subject might potentially respond.  
Assumption~\ref{assump:unconf} states that it may be possible to identify all pre-treatment covariates that are predictors of treatment or outcome. If $X$ contains all confounders, among subjects who share the same $X$ there is no association between $A$ and potential outcomes. 
A common practice, in reality, is to collect a large number of possible confounders in order to mitigate the violation of this assumption. However, including all available covariates in the analysis may introduce bias and variance of the causal effect estimator \citep{pearl2011invited,myers2011effects,brookhart2006variable,yang2020doubly}. Variable selection is hence an important procedure when estimating ATE. 
Assumption~\ref{assump:overlap} adds a restriction on the joint distribution of treatment assignment and covariates. Overlap is an important issue in estimating treatment effects from non-randomized trials. It describes the extent to which the range of data is the same across the two treatment groups. 
In fact, lack of overlap may affect all types of estimators. For matching estimators, that means it is difficult to find good matches; for weighting estimators, small overlap can result in extremely large weights; and for regression estimators, they may heavily rely on extrapolation. When Assumption~\ref{assump:overlap} is violated, a common practice is to trim the sample to restrict inference to the one with sufficient overlap \citep{yang2018asymptotic}, or to coarsen PS \citep{zhou2015coarsened}. 

Given the above stated assumptions, the ATE can be identified from observed data by conditioning on available covariates 
\begin{align}\label{eq:ate}
    \tau = E\{Y(1)\} - E\{Y(0)\} = E\{E(Y|A=1,X) - E(Y|A=0,X)\}
\end{align}
where the outer expectation is with respect to distribution of $X$ over entire population. 

In this section, we first review two kinds of \textit{singly robust} estimators based on either outcome modeling or treatment modeling, respectively in the sense that the consistency of the estimators relies on the correctness of the underlying model. Then we review various \textit{doubly robust} estimators that combine outcome modeling and treatment modeling in estimating ATE. 
Augmented inverse probability treatment weighting \citep[AIPTW,][]{lunceford2004stratification,glynn2010introduction,cao2009improving,robins2000marginal} belongs to a class of weighting estimators. AIPTW is a combination of the basic inverse probability weighting estimator and a weighted average of the outcome imputation estimators. AIPTW improves the IPW estimator by fully utilizing information about both treatment and outcome. 
Targeted maximum likelihood estimation \citep[TMLE,][]{van2006targeted,van2011targeted} is a regression estimator based on maximum likelihood estimation and includes a ``targeting" step that optimizes the bias-variance tradeoff for the causal estimand. 
Double score matching \citep[DSM,][]{yang2020multiply,zhang2021practical} belongs to the class of matching estimators. DSM matches on both propensity score and prognostic score. 
Penalized spline of propensity methods for treatment comparison \citep[PENCOMP,][]{zhang2009extensions,zhou2019penalized} is an example of doubly robust regression estimator. PENCOMP estimates causal effects by imputing missing potential outcomes with flexible spline models using multiple imputations. 

\subsection{Estimators for ATE Based on Outcome Modeling}

A traditional way to adjust for covariate imbalance in observational studies is via the formulation of a regression model for the outcome $Y$ on $A$ and $X$. That is, we can estimate the regression $E(Y|A,X)$ by modeling on the observed data. Given the stated assumptions Section~\ref{sec:assump}, the ATE can be identified by equation~\eqref{eq:ate}. 

An example of an regression imputation estimator can be obtained by fitting a linear regression of $Y$ given $A$ and all $X$, which is given by
\begin{align}\label{eq:ols}
    E(Y|A,X) = \alpha_0 + \alpha_A A + X^T \alpha_X.
\end{align}
Suppose this is indeed the true regression model, by equation~\eqref{eq:ate}, $\tau = \alpha_0 + \alpha_A \cdot 1 + X^T \alpha_X - (\alpha_0 + \alpha_A \cdot 0 + X^T \alpha_X) = \alpha_A.$ 
The ATE can be obtained directly from the coefficient for $A$, i.e. $\alpha_A$ in the regression model. 
If the true regression is specified, this estimator is consistent of $\tau$, that is, $\hat \alpha_A \overset{p}{\to} \tau$. Hence, this regression imputation estimator is a singly robust estimator in the sense that it is consistent when the outcome model is correctly specified. 

However, model~\eqref{eq:ols} assumes a constant treatment effect and could be severely biased in the case of heterogeneous treatment effects. In practice, treatment effects may vary across subjects. 
The regression imputation estimator is usually obtained by modeling outcome separately within each treatment arm rather than by using a single model~\eqref{eq:ols}. 
Note that the regression above can be made more general as a general parametric model, since $Y$ can be of any type. 
The missing potential outcomes are then imputed by their predictions from the corresponding posited models. The regression imputation estimator is given by the difference in the averages of potential outcomes. This can help to address heterogeneity in treatment effects, however, the issue of model misspecification still exists. Also, in case of a near violation of Assumption~\ref{assump:overlap}, the outcome model-based approaches rely on extrapolation.

\subsection{Estimators for ATE Based on Treatment Modeling}

Another class of ATE estimator for covariate adjustment relies on the treatment model, or the propensity score model. The propensity score is defined as the probability of treatment given covariates, i.e., $e(X) = E(A|X) = P(A=1 | X). $

Under the assumptions defined in Section~\ref{sec:assump}, given the propensity score, the potential outcomes and treatment assignment are independent \citep{rosenbaum1983central}, i.e., $\{Y(0), Y(1)\} \perp A ~|~ e(X)$. Traditionally, the estimation of propensity is by using a logistic regression where 
$e(X,\beta) = \logit^{-1} \{\exp(\beta_0+X^T\beta_X)\}$. 

Consider the inverse of the propensity score as a weight for the outcome, under the assumptions stated in Section~\ref{sec:assump} and the true propensity score
\begin{align*}
    E\Big\{\frac{ZY}{e(X)}\Big\} &= E\Big\{\frac{ZY(1)}{e(X)}\Big\} = E\Big[E\Big\{\frac{ZY(1)}{e(X)} \Big| Y(1), X\Big\}\Big] \\
    &= E\Big\{\frac{Y(1)}{e(X)} E(Z|Y(1), X)\Big\} = E\Big\{\frac{Y(1)}{e(X)} E(Z|X)\Big\} = E\Big\{\frac{Y(1)}{e(X)} e(X)\Big\} = E\{Y(1)\}. 
\end{align*}
Similarly, $E\{ (1-Z)Y/e(X) \} = E\{Y(0)\}$. 

A well-known common estimator based on propensity score is Inverse Probability Treatment Weighting (IPTW) \citep{lunceford2004stratification}. Specifically, IPTW estimates $\tau$ by the difference of inverse probability of treatment weighted averages, which is given by
\begin{align*}
    \hat\tau_{IPTW} = \frac{1}{n}\sum_{i=1}^n \frac{A_iY_i}{\hat e(X_i)} - \frac{1}{n}\sum_{i=1}^n \frac{(1-A_i)Y_i}{1-\hat e(X_i)}. 
\end{align*}
The inverse weights create a pseudo-population where there is no confounding so the weighted averages can reflect averages in the target population. If the true model for the propensity score model is specified, $\hat\tau_{IPTW}$ is a consistent estimator of $\tau$. Hence, IPTW is a singly robust estimator in the sense that it is consistent when the treatment model is correctly specified. A major drawback of the IPTW estimator is that IPTW is highly unstable due to weighting by the inverse of the propensity score. If any propensity score is close to zero or one, the IPTW estimator may be extreme. 

Variable selection is an important consideration when constructing the propensity score. 
Including predictors of treatment but not outcome, i.e., instrument variables, in the treatment model or the outcome model may amplify bias and variance of the causal estimator \citep{pearl2011invited,myers2011effects}. 
Including outcome predictors, on the other hand, could boost efficiency \citep{brookhart2006variable,yang2020doubly,pmlr-v177-tan22a}. 
Therefore, variable selection is needed before the estimation of treatment effects to remove variables not related to outcomes. 
Besides, weighting estimators are inferior in the case of extreme propensity scores \citep{kang2007demystifying}. Poor overlap in propensity score distributions can result in extremely large weights, leading to an unstable estimator with a large variance. 
Furthermore, model misspecification of the propensity score would also lead to a biased causal estimate.

\subsection{Augmented Inverse Probability Treatment Weighted (AIPTW)}

AIPTW estimator is a weighting based estimator that improves IPTW by fully utilizing information about both the treatment assignment and the outcome \citep{glynn2010introduction}. It is a combination of IPTW estimator and a weighted average of the outcome imputation estimators. Specifically, AIPTW is given by 
\begin{align}\label{eq:aiptw}
    \hat\tau_{AIPTW} = \frac{1}{n}\sum_{i=1}^n
    \Bigg\{
    \Big\{
    \frac{A_iY_i}{\hat e(X_i)} -  \frac{(1-A_i)Y_i}{1-\hat e(X_i)} 
    \Big\} - \frac{A_i - \hat e(X_i)}{\hat e(X_i)\{1-\hat e(X_i)\}} \Big[\{1-\hat e(X_i)\} \hat m_1(X_i) + \hat e(X_i) \hat m_0(X_i)\Big]
    \Bigg\}
\end{align}
where $m_1(X)$ is a postulated model for $E(Y|A=1,X)$ and $m_0(X)$ is a postulated model for $E(Y|A=0,X)$. 
The first line of equation~\eqref{eq:aiptw} is the same as $\hat\tau_{IPTW}$ and the rest adjusts this estimator by a weighted average of the two outcome imputation estimators. 
Rearranging terms in equation~\eqref{eq:aiptw}, $\hat\tau_{AIPTW}$ can be given by
\begin{align*}
    \hat\tau_{AIPTW} &= 
    \frac{1}{n}\sum_{i=1}^n
    \Big\{
    \frac{A_iY_i}{\hat e(X_i)}
    - \frac{A_i-\hat e(X_i)}{\hat e(X_i)} \hat m_1(X_i) \Big\}
    - \frac{1}{n}\sum_{i=1}^n
    \Big\{
    \frac{(1-A_i)Y_i}{1-\hat e(X_i)} 
    + \frac{A_i-\hat e(X_i)}{1-\hat e(X_i)} \hat m_0(X_i) 
    \Big\}
\end{align*}

AIPTW is a doubly robust estimator in that as long as either the outcome model is correct or the propensity score model is correct, $\hat\tau_{AIPTW}$ is a consistent estimator for $\tau$ \citep{glynn2010introduction}. Also, it enjoys good large-sample theoretical properties that it can be shown to be asymptotically normally distributed via derivation through the theory of $M$-estimation. 
Bootstrap can also be used to obtain the standard error estimates \citep{imbens2004nonparametric}.  These standard error estimators tend to be reasonable unless the estimated propensity scores are very extreme as weighting estimators are inferior in the case of extreme propensity scores \citep{kang2007demystifying}. In those scenarios, AIPTW is less robust to data sparsity and near violation of the positivity assumption \citep{glynn2010introduction}.

Recently, machine learning has gained popularity in the field of causal inference \citep{peters2017elements,prosperi2020causal,pmlr-v162-tan22a,pittir43544,tan2022rise}. \cite{10.1111/ectj.12097} shows that the regression imputation and IPTW estimators using machine learning nuisance function estimators tend to have large finite sample biases. AIPTW, derived from the semiparametric efficiency theory, on the other hand, enjoys the \textit{rate double robustness} when combined with machine learning \citep{kang2007demystifying,10.1111/ectj.12097}. 
That is, these doubly robust estimators remain root-$n$ consistent and asymptotically normal when using machine learning approaches to estimating the nuisance functions. 
Incorporating AIPTW with machine learning could help to mitigate the impact of regularization bias and overfitting on causal estimate \citep{10.1111/ectj.12097}.

\subsection{Targeted Maximum Likelihood Estimation (TMLE)}

TMLE, introduced by \cite{van2006targeted}, is a maximum likelihood-based estimator that incorporates a ``targeting" step that optimizes the bias-variance tradeoff for the targeted estimator, i.e., ATE. Specifically, TMLE obtains initial outcome estimates via outcome modeling and propensity scores via treatment modeling, respectively. These initial outcome estimates are then updated to reduce the bias of confounding, which generates the so-call ``targeted" predicted outcome values. 

The estimation steps of TMLE are given as follows. First, initial outcome estimates are constructed by $\hat Y_1 = \hat E(Y|A=1,X)$ and $\hat Y_0 = \hat E(Y|A=0,X)$, respectively, and the propensity scores $\hat e(X)$ are estimated through treatment modeling. Then the targeted steps begin by first calculating the inverse propensity $H_a$ for each subject, 
\begin{align*}
    H_1(A=1,X) = \{\hat e(X)\}^{-1} \text{~~and~~} H_0(A=0,X) = -\{1-\hat e(X)\}^{-1}. 
\end{align*}
This is similar in form to inverse probability weights. Then, for the treatment arm and the control arm, separately, the observed outcome $Y$ is regressed on those estimated inverse propensity with fixed intercepts. Take a binary outcome as an example. A logistic transform can be applied with a binary outcome $Y$ where 
\begin{align*}
    logit\{E(Y|A=1,X)\} = logit(\hat Y_1) + \epsilon_1 H_1 \text{~~and~~}
    logit\{E(Y|A=0,X)\} = logit(\hat Y_0) + \epsilon_0 H_0. 
\end{align*}
In this way, we are able to generate updated, or so-called “targeted” estimates of the set of potential outcomes, incorporating information from the treatment mechanism in order to reduce the bias. 
The predicted outcomes are then updated to be 
\begin{align*}
    logit(\hat Y_1^*) = logit(\hat Y_1) + \hat\epsilon_1 H_1  \text{~~and~~} 
    logit(\hat Y_0^*) = logit(\hat Y_0) + \hat\epsilon_0 H_0
\end{align*}
The final estimates is given by calculating ATE as mean difference in targeted predicted outcome pairs across individuals
\begin{align}\label{eq:tmle}
    \hat\tau_{TMLE} = \frac{1}{n} \sum_{i=1}^n (\hat Y_{1i}^* - \hat Y_{0i}^*).
\end{align}

The variance of the TMLE estimator is obtained based on the efficient influence curve \citep{porter2011relative,van2006targeted,munoz2011super,van2007super}. In general, TMLE and AIPTW are both efficient and have the minimum asymptotic variance under the large-sample theory. However, it has been shown that under finite sample size or challenging scenarios such as misspecified models, and nearly violated positivity, TMLE may still provide causal estimates in the range of ATE since $\hat Y_a^*$ in equation~\eqref{eq:tmle} are range-preserving while AIPTW may not \citep{van2011targeted,porter2011relative,pirracchio2015improving}. The estimation of TMLE is usually coupled with SuperLearner \citep{munoz2011super,van2007super} for $\hat Y_a$ and $\hat e(X)$, which is an ensemble of multiple statistical and machine learning models. It learns an optimal weighted average of those models by giving higher weights to more accurate models, and has been proven to have high accuracy \citep{munoz2011super,van2007super}. The performance of TMLE continues to boost with the help of SuperLearner. Note that hybrid estimators have been proposed to resemble TMLE and AIPTW that use coarsened propensity score estimates instead of model-based ones \citep{zhou2015coarsened}. They may have better performance in case of severe model misspecification. However, the choice of coarsening mechanism and the coarsening parameters may introduce extra challenges \citep{zhou2015coarsened}, and thus these hybrid estimators are not included for comparison in the later simulation studies. 



\subsection{Double Score Matching (DSM)}

Matching methods are powerful because they can be used to re-create a randomized trial that is hidden under the observational study. 
DSM is a matching-based estimator that uses the double balancing properties of propensity score $e(X)$ and prognostic score $\Phi(X)$, the latter obtained via outcome modeling, before the matching is conducted \citep{yang2020multiply}. The prognostic score is defined as a balancing score where $\{Y(0),Y(1)\} \perp A | \Phi(X)$ \citep{hansen2008prognostic}. The combination of $e(X)$ and $\Phi(X)$, the double score, is also shown to be a balancing score \citep{antonelli2018doubly}. That is, $\{Y(0),Y(1)\} \perp A | \{e(X), \Phi(X)\}$. 

DSM estimator enjoys the double robustness property in that this result holds even if only one score is correctly specified. For unit $i$, the potential outcome under $A_i$ is the observed outcome $Y_i$. The potential outcome under $1 - A_i$ is not observed but can be imputed by the observed outcomes of the nearest $M$ units with $1-A_i$. 
Denote the augmented score $S=\{e(X),\Phi(X)^T\}^T$ as the matching variable,  $J_{S,i}$ as the index set for these matched subjects for subject $i$, and $K_{S,i} = \sum_{j=1}^n I(i \in J_{S,i})$ as the number of times subject $i$ is used as a match. The matched set is constructed with distance metric such as Mahalanobis distance on $S$ that combines propensity score and prognostic score. 
The initial DSM estimator of $\tau$ is
\begin{align*}
    \hat\tau_{DSM}^{(0)} = \frac{1}{n} \sum_{i=1}^n (2A_i - 1)(1+M^{-1} K_{S,i}) Y_i. 
\end{align*}
A de-biasing DSM estimator $\hat\tau_{DSM}$ suggested by \cite{yang2020multiply} further corrects bias by the method of sieves. Correctly the bias may help to improve finite sample performance in practice although this bias is asymptotically negligible \citep{yang2020multiply}. A wild bootstrap procedure is used to obtain the confidence interval based on \cite{otsu2017bootstrap} for matching estimators \citep{yang2020multiply}. 

Matching methods tend to be more stable tools when the propensity score is extreme \citep{stuart2010matching}. Matching estimators are robust to model misspecifications if the misspecified model belongs to the class of covariate scores \citep{waernbaum2012model}. 
DSM is robust against model misspecification of either the propensity score model or the prognostic score model \citep{antonelli2018doubly,yang2020multiply}. 
Specifically, DSM provides multiple protections to model misspecification by positing multiple candidate models for both propensity scores and prognostic scores. This helps DSM to achieve a near nominal coverage even under model misspecification \citep{yang2020multiply}. Furthermore, DSM can serve as a dimensional reduction tool in high-dimensional confounding. However, adding too many covariates could result in potential bias as matching estimators may not work well on high-dimensional covariates \citep{abadie2006large}. It is also pointed out that the number of posited models and their functional forms affect the efficiency of DSM in a complex way, resulting in an unstable performance if there is a large number of working models \citep{yang2020multiply,zhao2021outcome}. 
Hence, variable selection in matching estimators is needed to help identify outcome predictors for better efficiency and remove bias in estimating the ATE. 
\cite{zhang2021practical} investigate the performance of DSM under
different strategies of variable selection, using a caliper, and matching with or without replacement, providing the best practice. 
Also, as the success of matching depends on the functional forms of posited models, flexible machine learning methods can be adopted in the modeling of propensity scores and prognostic scores before matching.


\subsection{Penalized Spline of Propensity Methods for Treatment Comparison (PENCOMP)}

PENCOMP is a type of regression- and multiple imputation-based approach \citep{zhou2019penalized}. It builds on the method of the penalized spline of propensity prediction previously used in missing data problems \citep{little2004robust,zhang2009extensions}. Specifically, PENCOMP obtains propensity score $e(X)$ via treatment modeling and uses spline-based regressions with propensity score included for outcome modeling.  
Under the assumptions in Section~\ref{sec:assump}, PENCOMP has a double robustness property for estimating ATE. 

PENCOMP uses Rubin’s rules for combining multiply
imputed datasets. The method first generates a bootstrap sample $b$ from the original data stratified on the treatment group. The propensity score $e(X)$ is then estimated. Then the potential outcomes for the treatments not assigned to subjects are predicted with regression models that include splines on the logit of the propensity to be assigned that treatment, plus other outcome predictors. For each treatment group, each regression model is fitted separately. Specifically, the regression model fitted for subjects with $A_i=a$ is given by
\begin{align*}
    E\{Y_i(a) | X_i,A_i=a\} = s\{\hat e(X_i)\} + g\{\hat e(X_i), X_i\}, ~~ i \in \{ i: A_i=a \}
\end{align*}
where $s\{\hat e(X)\}$ is a penalized spline with pre-specified knots, and $g\{\hat e(X), X\}$ is a parametric function of outcome predictors as well as the estimated propensity. The missing potential outcome of a subject is then imputed based on the predictive distribution of $E\{Y(a) | X,A\}$. The bootstrap estimate $\hat\tau_{PENCOMP}^{(b)}$ is the difference in the treatment means based on the observed outcome and the imputed values of $Y$. 
The above procedure is repeated multiple ($B$) times. The final ATE estimate $\hat\tau_{PENCOMP} = B^{-1}\sum_{b=1}^B \hat\tau_{PENCOMP}^{(b)}$. The confidence interval of $\hat\tau_{PENCOMP}$ is generated from this procedure. 

According to \cite{zhou2019penalized}, PENCOMP achieves comparable performance with AIPTW in terms of bias, RMSE, and coverage under settings of low confounding and correctly specified model with linear settings. 
Also, PENCOMP has some advantages in nonlinear settings compared to AIPTW \citep{zhou2019penalized}. 
However, in the case of model misspecification, there exists a severe overcoverage with wider confidence interval for PENCOMP compared to AIPTW even under a linear setting \cite{zhou2019penalized}. It is also pointed out by \cite{kang2007demystifying} that if regression models are misspecified, doubly robust methods could suffer from larger bias compared to singly robust methods. 
Besides, the choice of the splines and knots can be challenging in practice. 




\section{Simulation Studies}
\label{s:sim}

In this section, we design Monte Carlo simulations to compare each doubly robust method under different scenarios mimicking real-world data where there is high nonlinearity in the relationship between covariates and treatment, and the relationship between covariates and outcome. Specifically, we consider settings with complex data generative models with multivariate covariates. We also consider cases of different degrees of separation of the propensity score distributions where the propensity scores may be close to zero or one. 
In Table~\ref{tab:code}, we provide the open-source software or code that implements the surveyed DR estimators. The code of our simulation studies can be found on GitHub (\texttt{https://github.com/ellenxtan/RealWorld-DoublyRobustML}).

\begin{table}[htp]
\centering
\caption{Open-source software or code that implements the surveyed DR estimators.}
\label{tab:code}
\begin{tabular}{@{}ll@{}}
\toprule
DR estimators & Open source code or package \\ \midrule
AIPTW & \texttt{AIPW} \citep{zhong_aipw_2021}  \\
TMLE & \texttt{tmle} \citep{gruber2012tmle} \\
DSM & \texttt{dsmatch} \citep{yang2020doubly} \\
PENCOMP & \texttt{PENCOMP} \citep{zhou2019penalized} \\ \bottomrule
\end{tabular}
\end{table}

\subsection{Data Generating Process}

\begin{figure}[htp]
 \centerline{\includegraphics[width=0.55\linewidth]{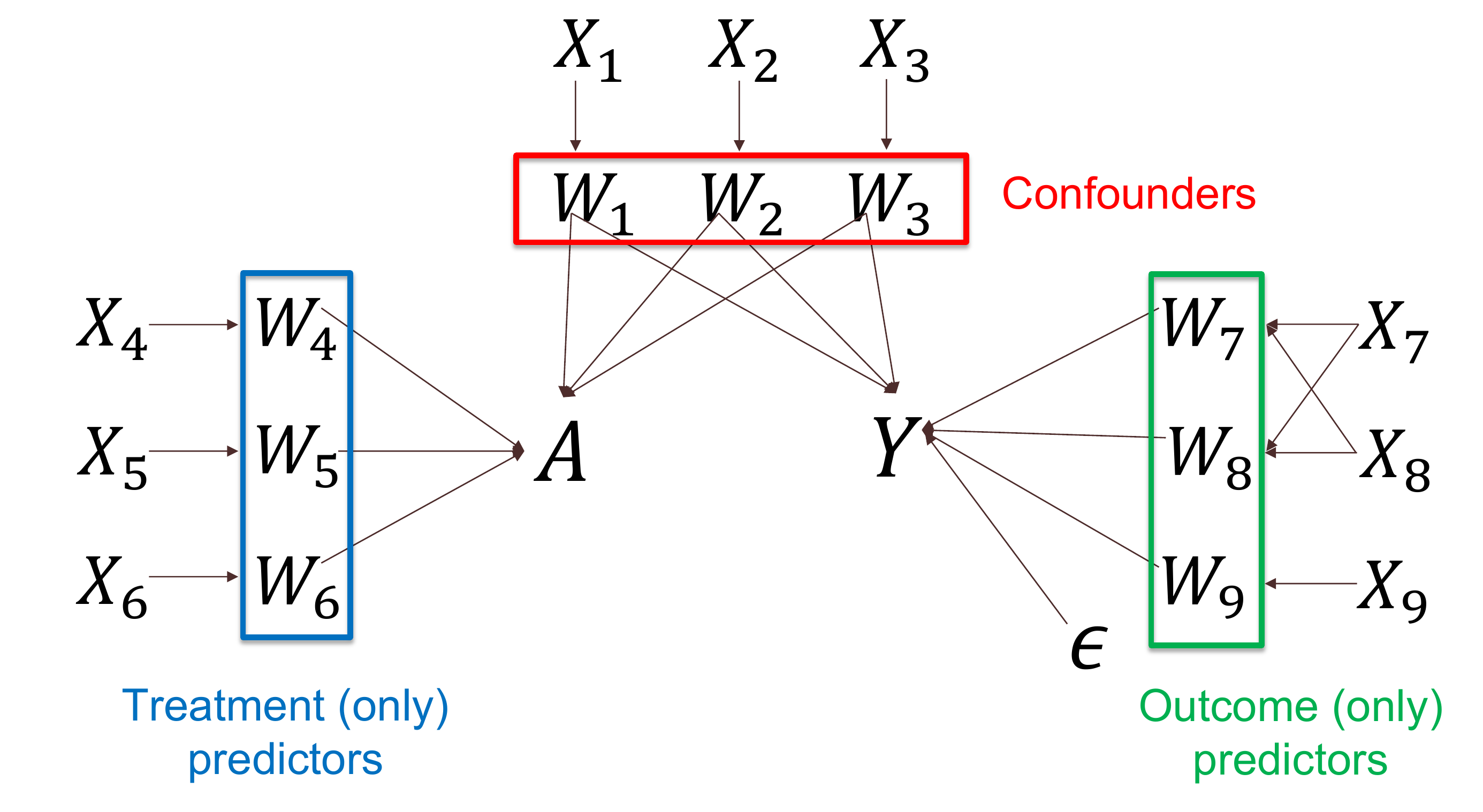}}
 \caption{Illustration of variables involved in simulation studies.}
 \label{f:dag}
\end{figure}

We design the data generating procedure following the work of \cite{leacy2014joint}, but with the nonlinearity of treatment and outcome surfaces considered. The sample size is set to be $n=2000$ throughout. 
The relationship of variables generated in simulation is illustrated in Figure~\ref{f:dag}. 
First of all, we generate $m=9$ independent and identically distributed standard normal variables $X \in R^9 \sim N(0,1)$. Let $W \in R^9$ be a nonlinear transformation of $X$ where $W_1, W_2, W_3$ are confounders (i.e., predictors of both treatment assignment and outcome), $W_4, W_5, W_6$ are treatment (only) predictors or instrumental variables, $W_7, W_8, W_9$ are outcome (only) predictors. Specifically, we let $W_1 = \exp(X_1 / 2),$ $W_2 = \exp(X_2 / 3),$ $W_3 = X_3^2,$ $W_4 = X_4^2,$ $W_5 = X_5,$ $W_6 = X_6,$ $W_7 = X_7 + X_8,$ $W_8 = X_7^2 + X_8^2,$ $W_9 = X_9^3$. All $W$ are standardized to have mean zero and variance 1. 

We generate the binary treatment indicator $A$ following Bernoulli$\{e(X)\}$ where $e(X)$ is the propensity score model. The propensity score model is designed to be a linear combination of confounders and treatment predictors in terms of $W$. 
To assess the property of different degrees on the separation of propensity score distributions, we consider two settings for the PS overlap to test the performance of estimators. A large overlap is an ideal case, which means there is a reasonable amount of common support for the treated subjects and the control subjects. The case of a small overlap, on the other hand, indicates a majority of the treated subjects may fail to find any suitable control in their neighborhoods. This could be common in scenarios such as rare diseases in practice. Specifically, we design the large overlap case to be 
$logit^{-1}\{e(X)\} = (-3 -W_1 + 2W_2 - 3W_3 + 3W_4 + 2W_5 + W_6) / 15$. 
We design the case of small overlap to be 
$logit^{-1}\{e(X)\} = (-8W_1 + 1.5W_2 + 0.5W_3 - 0.5W_4 + 2.5W_5 - 0.5W_6) / 5$. 

We generate a continuous outcome, which is a linear combination of confounders and outcome predictors in terms of W. Specifically, we let $Y(0) = -2 +1.5W_1 - 2W_2 + 1.5W_3 + 2.5W_7 - W_8 + W_9 + \epsilon$ where $\epsilon \sim N(0,1).$ We assume that only $X$'s are included in the candidate set therefore nonlinearity in the outcome and treatment models induced by $W$'s should be explicitly modeled by the analyst. 
This mimics that in practice there exists potential nonlinearity between covariates and outcome and this nonlinearity is needed to be considered when fitting models. We consider both a homogeneous treatment effect setting and a heterogeneous treatment effect setting. For the homogeneous setting, we consider a constant treatment effect. We use $\tau = 0$ in our simulation. Specifically, 
$Y(1) = Y(0) + \tau$. For the heterogeneous setting, we allow the treatment effect to vary by some covariates, and here $W_1$ and $W_3$. Specifically, $Y(1) = Y(0) + \tau + 5W_1 + 3W_3 + 2W_1 W_3$. We try $\tau=0$. 
The Observed outcome $Y$ is given by $Y = Y(1)  A + Y(0)  (1 - A)$.

To summarize, there are four scenarios in total, which are
\begin{enumerate}[label={\arabic*)}] 
    \item homogeneous treatment effect with a large PS overlap (Homo TE + large overlap)
    \item homogeneous treatment effect with a small PS overlap (Homo TE + small overlap)
    \item heterogeneous treatment effect with a large PS overlap (Hetero TE + large overlap)
    \item heterogeneous treatment effect with a small PS overlap (Hetero TE + small overlap)
\end{enumerate}

\subsection{Evaluation Metrics}

We consider the following metrics averaged over observations and simulation replications.
\begin{itemize}
    \item Bias: $bias = R^{-1}\sum_{r=1}^R \{\hat{\tau}^{(r)} - \tau\}$
    \item MSE: $mse = R^{-1}\sum_{r=1}^R \{\hat{\tau}^{(r)} - \tau\}^2$
    \item Confidence interval coverage: $coverage = R^{-1}\sum_{r=1}^R \mathbbm{1}\{\tau \in (\hat{\tau}^{(r)}_{L,0.05}, \hat{\tau}^{(r)}_{U,0.05})\}$
    \item Confidence interval width: $width = R^{-1}\sum_{r=1}^R \{\hat{\tau}^{(r)}_{U,0.05} - \hat{\tau}^{(r)}_{L,0.05}\}$
    \item Type I error: $\alpha = 1 - R^{-1}\sum_{r=1}^R \mathbbm{1}\{0 \in (\hat{\tau}^{(r)}_{L,0.05}, \hat{\tau}^{(r)}_{U,0.05})\}$
    \item Variance ratio: $var.ratio = R^{-1}\sum_{r=1}^R [var_{m}\{\hat{\tau}^{(r)}\}] \{var_{b}(\hat{\tau})\}^{-1}$ 
    where $var_{m}(\hat{\tau}^{(r)})$ is squared estimated standard error $\hat{\tau}^{(r)}$ for the $r$-th replication and $var_{b}(\hat{\tau})$ is variance of $\hat{\tau}$ over $R$ replications. Variance ratio measures the ratio between mean variance of an estimator over $R$ replicates and variance of estimates from the $R$ replicates. It evaluates the performance of model-based standard errors of $\hat{\tau}$ by comparing them with simulation variance reflecting the true variability of estimated $\tau$. 
\end{itemize}

\subsection{Analysis Steps and Compared Estimators}
\label{sec:analysis-step}

For analysis steps, we generate simulated data of $n=2000$ subjects. 
We apply the Lasso \citep{tibshirani1996regression}, a variable selection technique before the estimation in order to remove variables that are not related to outcomes. The outcome predictors are chosen by using all $X$'s as covariates and the observed outcomes as response. Five-fold cross-validation is used to select the best tuning parameter in Lasso, with the cross-validation deviance within 1 standard error of the minimum, as recommended by \cite{zhang2021practical}. 
Outcome predictors are obtained for each treatment arm respectively. 
We then use GLM, GAM, or SuperLearner separately to model PS and/or outcome before estimating the final ATE. 
The candidate learners in the SuperLearner library are: linear regression, stepwise regression, GAM, and Bayesian Additive Regression Trees \citep{chipman2010bart}. 
Doubly robust estimators for comparison are AIPTW, TMLE, DSM, and PENCOMP. We also include for comparison the singly robust estimator IPTW, and a regression imputation estimator, denoted as IMP, which fits a twin outcome model on covariates separated by the treatment group and the control group and then imputes the missing potential outcomes with the posited models. 
Note that IPTW involves fitting the PS model only. 
We generate $R=1000$ replications of simulated data sets. For AIPTW and TMLE, bootstrap CIs are also computed. Each replication uses $B=500$ for the bootstrap CI.  


\begin{figure}[htp]
\centering
 \begin{subfigure}{0.9\textwidth}
  \centerline{\includegraphics[width=\linewidth]{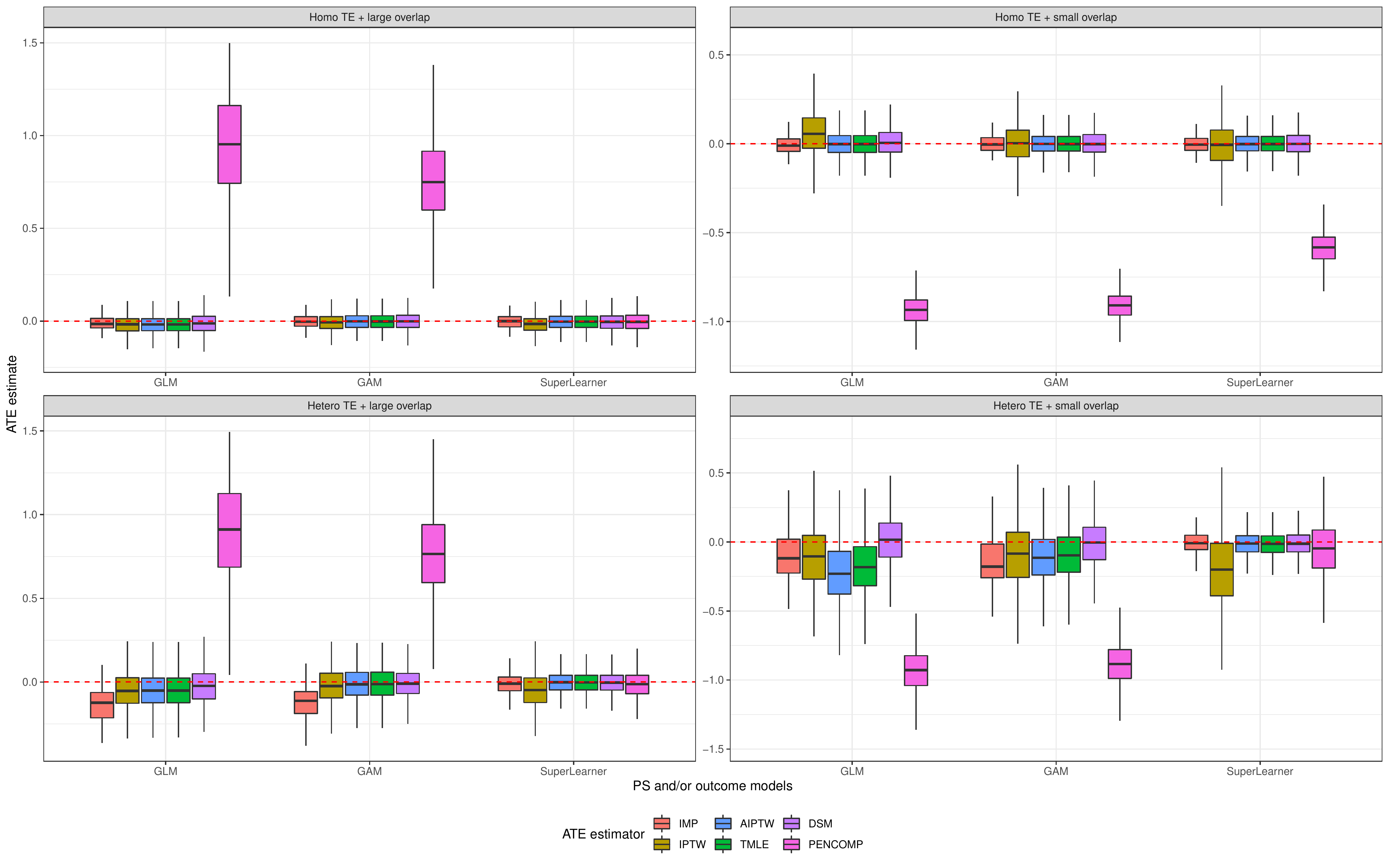}}
  \caption{}
 \end{subfigure}
 \begin{subfigure}{0.9\textwidth}
  \centerline{\includegraphics[width=\linewidth]{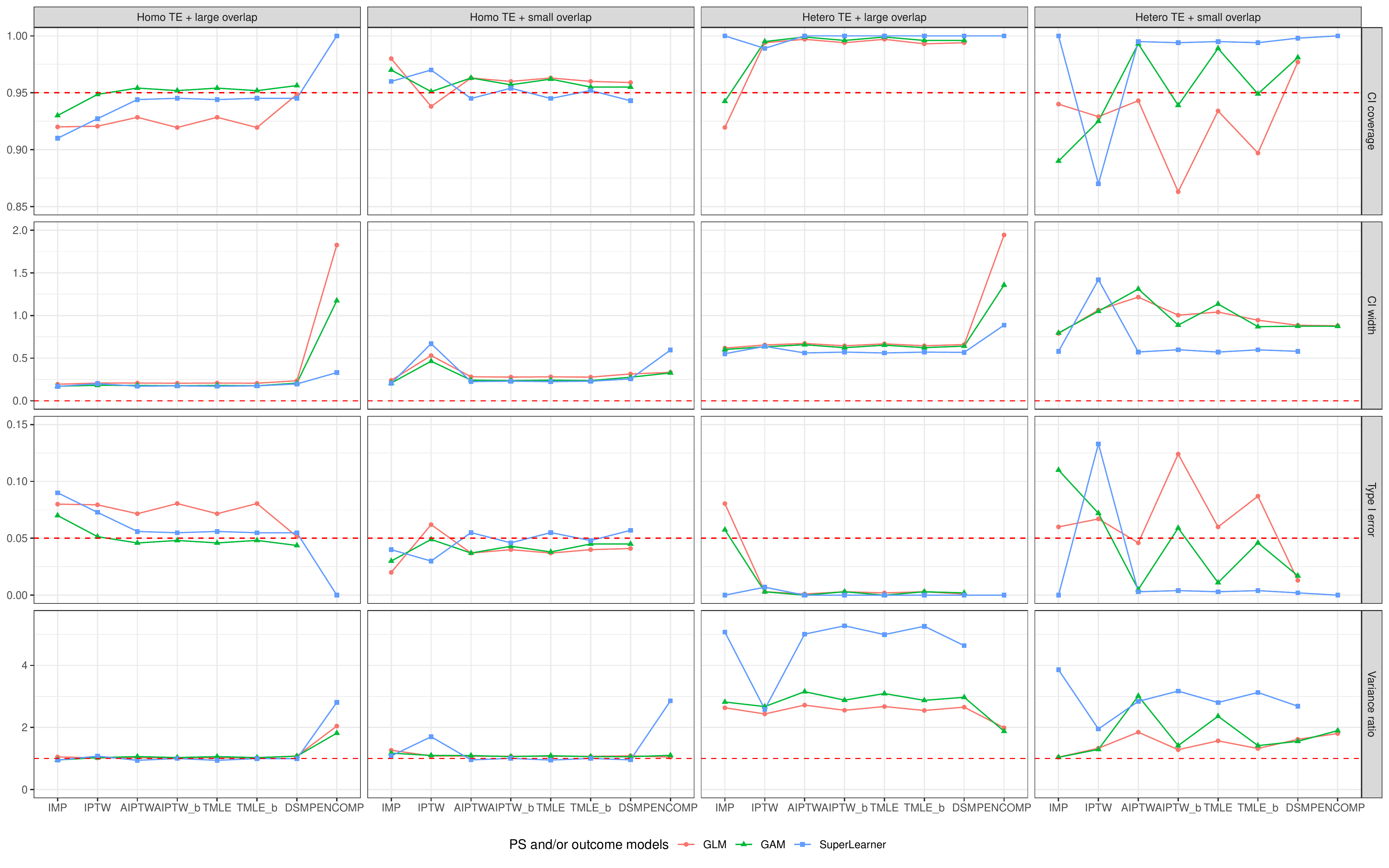}}
  \caption{}
 \end{subfigure}
 \caption{
Performance of various causal estimators under the four simulation scenarios. 
(a) Box plots of the estimators. 
The red dotted lines indicate the ground truth ATE. 
(b) Performance of the estimators in terms of CI coverage and width, type I error, and variance ratio. 
The red dotted lines indicate the ideal value or threshold of corresponding metrics. 
}
 \label{f:pure_box_line}
\end{figure}


\begin{sidewaysfigure}
    \centerline{\includegraphics[width=\linewidth]{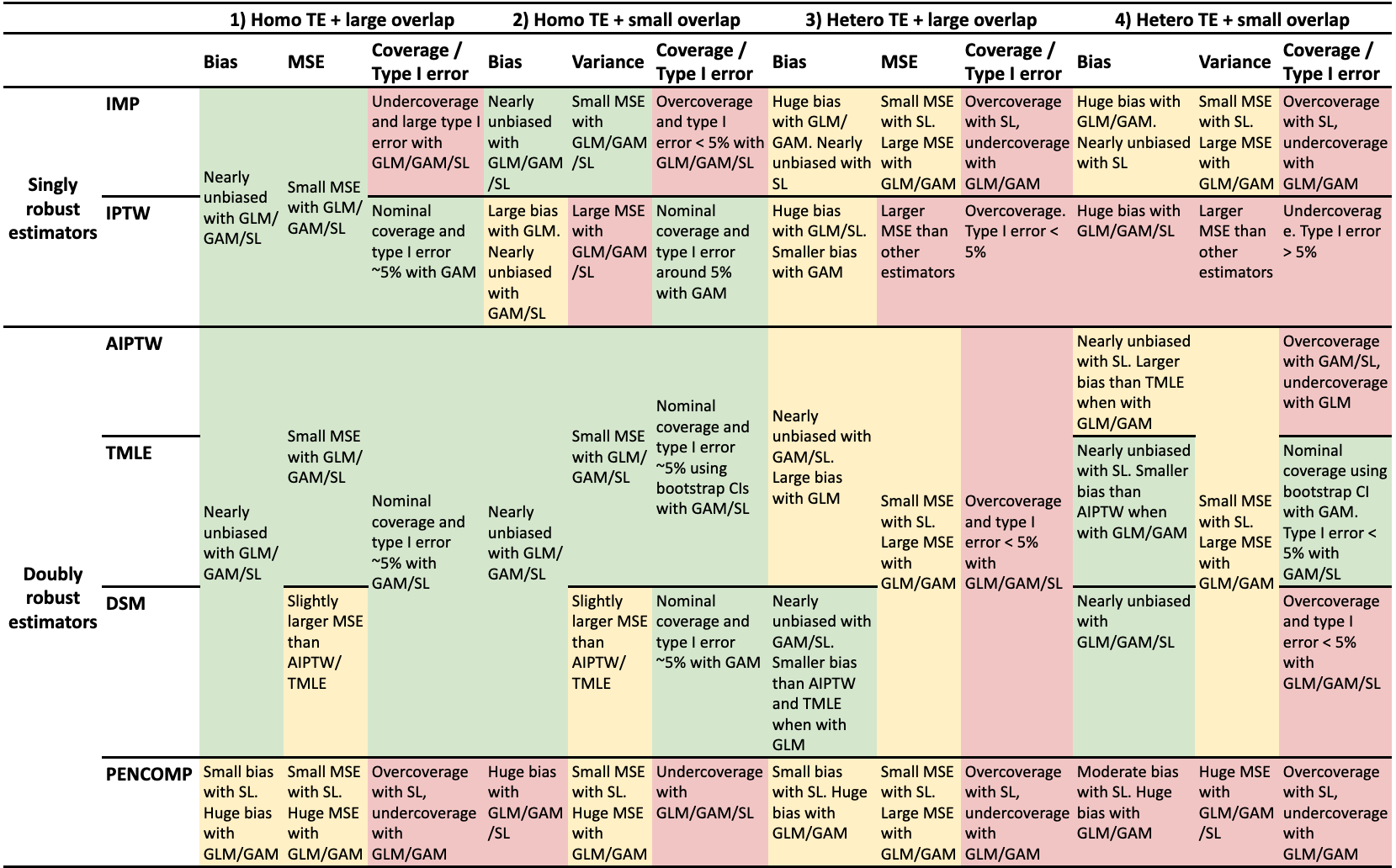}}
 \caption{Summary of the performance of various causal estimators under the four simulation scenarios. Colors indicate the performance of the estimator with green, yellow, and red meaning good, average, and poor performance, respectively.}
 \label{f:sim_smry}
\end{sidewaysfigure}

\subsection{Simulation Results}

Figure~\ref{f:pure_box_line} shows the performance of various causal estimators under the four simulation scenarios. 
A summary of the performance of the estimators is given in Figure~\ref{f:sim_smry}. 
In general, doubly robust estimators outperform singly robust estimators. 
IMP tends to have robust performance by chance under homogeneous settings, but may fail under heterogeneous settings with GLM and GAM because of extrapolation and model misspecification. With SuperLearner, IMP greatly reduces bias and variance. 
IPTW has improved performance with GAM in terms of bias and MSE. However, with GLM and SuperLearner, IPTW may suffer from large bias and MSE. IPTW may achieve a small bias under homogeneous treatment effects but fails under settings of heterogeneous treatment effects. 
Compared to overlap between propensity score distribution, treatment effect heterogeneity has a larger effect on the performance of doubly robust estimators. 
Using SuperLearner for treatment and outcome modeling, doubly robust estimators achieve the smallest bias and MSE. Using GLM for treatment and outcome modeling could suffer from huge bias because the relationship between covariates and the outcome is nonlinear. Using GAM would help improve the bias and MSE (bias in particular), but some nonlinearity may still be hard to capture. 
Under the relatively easy homogeneous treatment effect setting, doubly robust estimators including AIPTW, TMLE, and DSM achieve a nominal confidence interval. Under challenging settings such as the heterogeneous treatment effect, all doubly robust estimators suffer from overcoverage. 
Overall, TMLE and AIPTW enjoy the most favorable performance with SuperLearner. They have minimal bias and MSE, especially with SuperLearner. 
Under the relatively simple homogeneous treatment effect setting, TMLE and AIPTW achieve a nominal coverage, and with SuperLearner, these two estimators even achieve the smallest CI width and control type I error under 5\%. 
Under the challenging settings, SuperLearner could inflate the variance ratio and show an overcoverage issue.  
With GAM, TMLE and AIPTW may be able to achieve a nominal confidence interval by using bootstrap (see in the setting of heterogeneous treatment effects with a small overlap). 
However, in terms of model misspecification, TMLE is more robust than AIPTW with GLM and GAM. TMLE tends to have a smaller bias and MSE. 
DSM is more robust in terms of bias in the case of model misspecification compared to other doubly robust estimators (typically revealed in settings of heterogeneous treatment effects). DSM with GAM may outperform DSM with SuperLearner in terms of bias and variance ratio (see in the setting of heterogeneous treatment effects with a large overlap). 
PENCOMP suffers from severe bias and MSE with GLM or GAM. PENCOMP has improved performance with SuperLearner in terms of bias and MSE. However, PENCOMP suffers from a severe overcoverage and a large type I error in nearly all settings.

\section{A Real-World Application}
\label{s:real}

\begin{figure}[htp]
 \centerline{\includegraphics[width=0.8\linewidth]{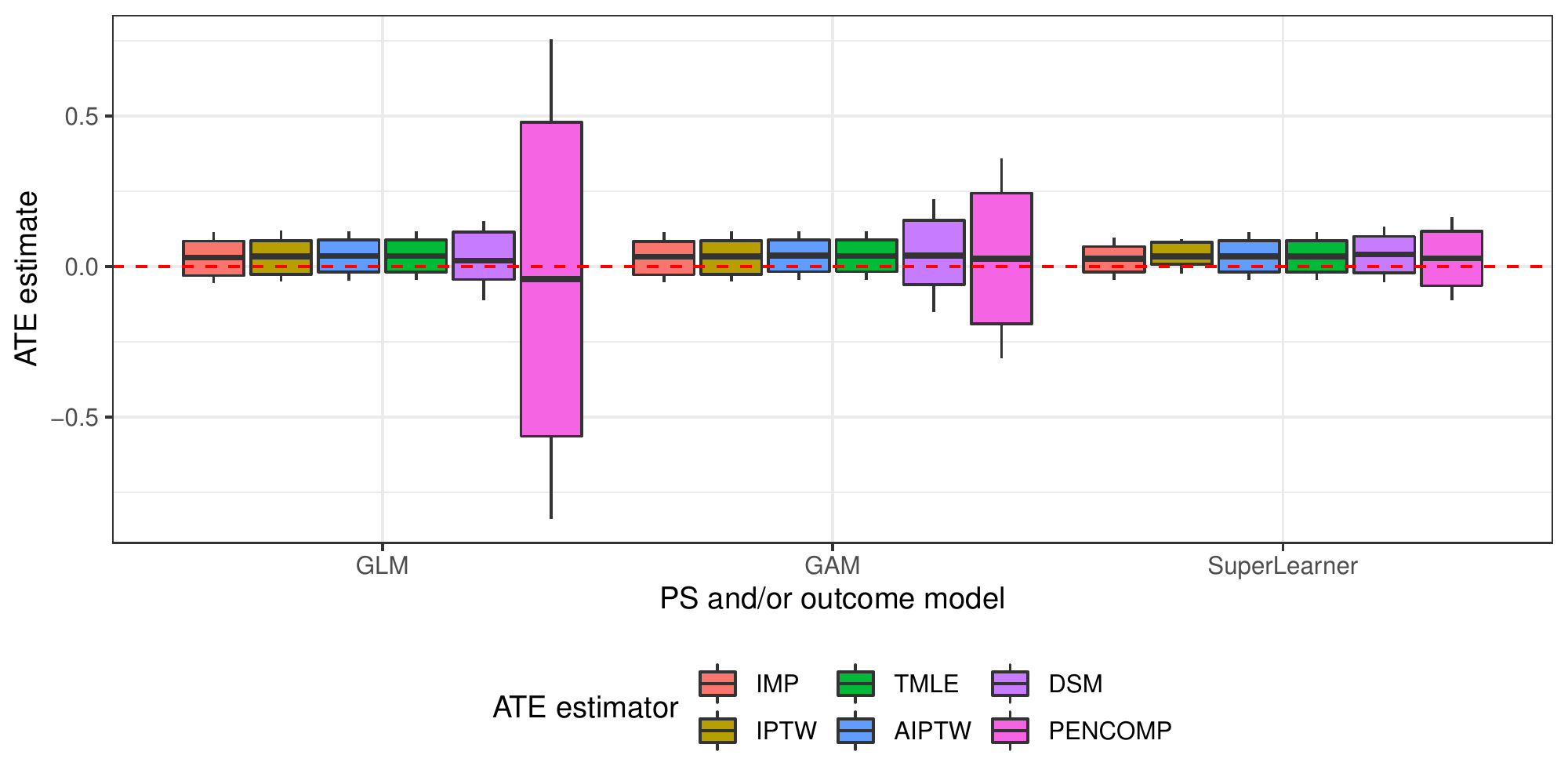}}
 \caption{Performance of various causal estimators on the real-data application. Different colors imply different causal estimators, x-axis differentiate the PS and/or outcome models. The red dotted line indicates a zero ATE.}
 \label{f:real_box}
\end{figure}

We apply different causal estimators to a real-world application, the Reflections study (REFL), which is a study of real-world examination of fibromyalgia for longitudinal evaluation of costs and treatments \citep{robinson2012burden}. 
We focus the analysis on opioid treatment arm (OPI cohort), and non-narcotic opioid-like treatment arm (TRA cohort). 
There are 544 patients in total. The outcome of interest is the change from baseline to LOCF in the total score of the Fibromyalgia Impact Questionnaire (FIQ), which is a continuous variable ranging from 0 to 80. There are 69 covariates in total, 24 of which are continuous variables and the other 45 are binary variables. Earlier studies showed there is no difference in FIQ among the treatment groups \citep{robinson2012burden, yang2016propensity}. Here we apply the compared causal estimators to estimate the causal effect of treatments for fibromyalgia on the FIQ score. The analysis steps are similar to those in Section~\ref{sec:analysis-step}. 
Figure~\ref{f:real_box} shows the performance of different estimators using different PS and outcome modeling. The estimated treatment effect is about 0.03 with confidence intervals of all doubly robust estimators including zero, which indicates there is no evidence that there are treatment effects between the OPI cohort and the TRA cohort on the FIQ score. With SuperLearner, the estimators achieve the smallest standard error.

\begin{figure}[!htp]
\centering
 \begin{subfigure}{0.9\textwidth}
  \centerline{\includegraphics[width=\linewidth]{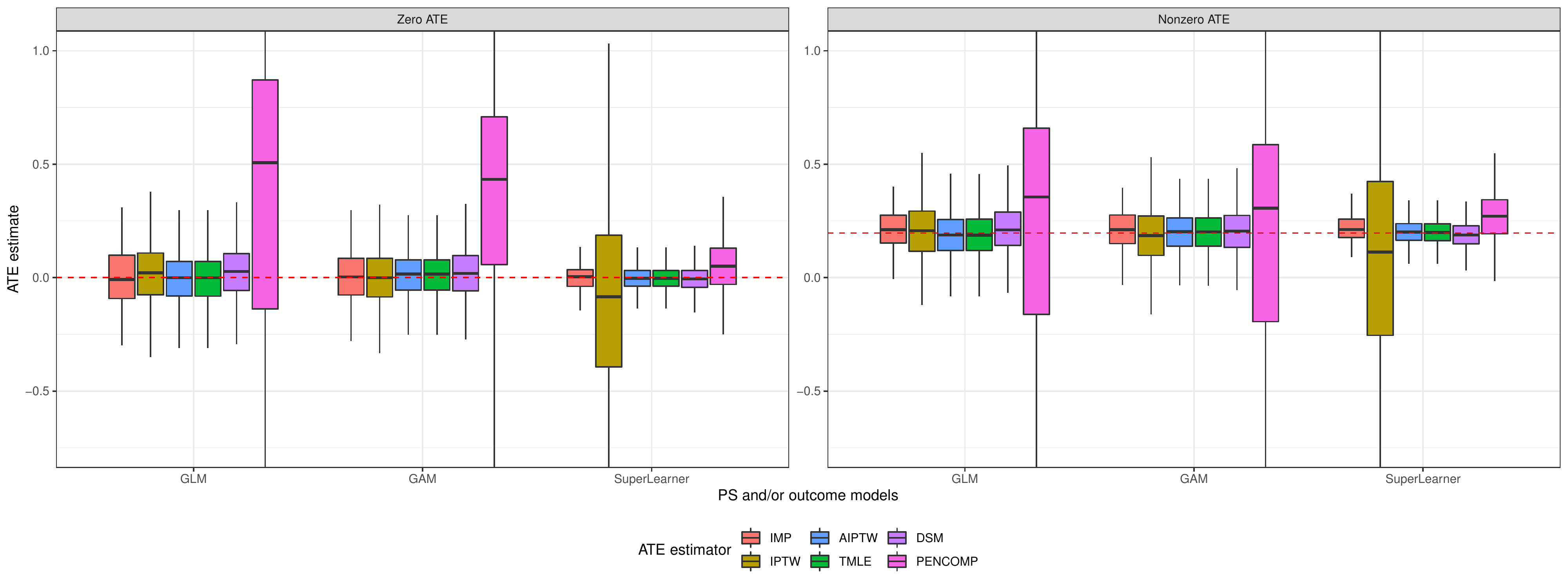}}
  \caption{}
 \end{subfigure}
 \begin{subfigure}{0.9\textwidth}
  \centerline{\includegraphics[width=\linewidth]{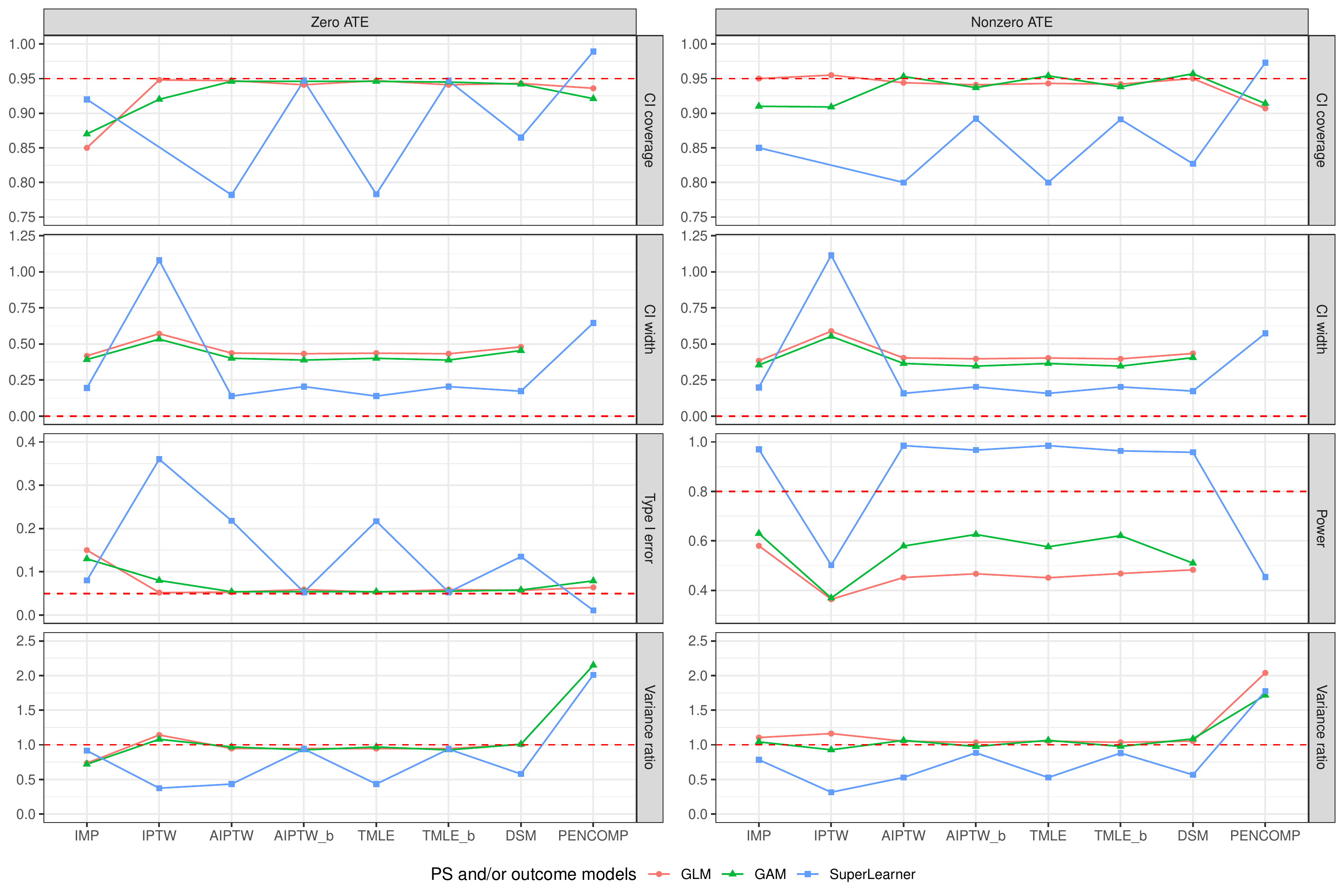}}
  \caption{}
 \end{subfigure}
 \caption{
Performance of various causal estimators on the simulated REFL study considering both a homogeneous treatment effect design (left, Zero ATE) and a heterogeneous treatment effect design (right, Nonzero ATE), respectively. 
(a) Box plots of the estimators. 
The red dotted lines indicate the ground truth ATE. 
(b) Performance of the estimators in terms of CI coverage and width, type I error, and variance ratio. 
The red dotted lines indicate the ideal value or threshold of corresponding metrics. 
 }
 \label{f:sim_refl_box_line}
\end{figure}

\subsection{Simulated REFL Study}

Motivated by the real-world REFL study, we are interested in evaluating the effects of a higher dimension of covariates and the effects of binary variables on estimating treatment effects. The simulated REFL study is hence designed to mimic the data distribution in the real REFL study. We generate covariates X using Iman-Conover transformation \citep{iman1982distribution} to simulate correlated covariates from the real REFL data. The simulated REFL data is of sample size 2000 (1000 for OPI cohort and 1000 for TRA cohort). 

To simulate the treatment indicator $A$, we first generate the treatment model by fitting an XGBoost \citep{chen2016xgboost} with cross-validation tuning using the real REFL data. This is referred to as the ``true" treatment model. Specifically, the treatment indicator is used as the outcome and all patient characteristics are included as covariates. We then generate the simulated treatment assignment $A$ for the simulated REFL data following Bernoulli($\hat{p}$) with $\hat{p}$ estimated from the ``true" treatment model. Similarly, the simulated outcomes $Y$ are obtained from the outcome model which is fit by an XGBoost with cross-validation tuning. This is referred to as the ``true" outcome model. 

Two simulation scenarios are designed with different ``true" outcome models. 
The first one is the zero treatment effect (Zero ATE) scenario, where the ``true" outcome model is based on fitting an XGBoost to outcome data with no treatment indicator, then using the predicted value from that model, denoted as $\hat{Y}$, to generate simulated outcomes by adding a Gaussian noise with variation obtained from the cross-validation process. The resulting simulated outcomes are considered ``observed" outcomes, and $\hat{Y}$'s are considered ``truth". 
The second one is the nonzero treatment effect (Nonzero ATE) scenario, where the ``true" outcome model is constructed by fitting an XGBoost to outcome data with $A$ simulated from the ``true" treatment model as a covariate and other covariates included in the candidate set. The corresponding predicted value $\hat{Y}$ from that fitted model is used to simulate data. 
There are 1000 simulated datasets generated, each one conducts a bootstrap 500 times. 

Figure~\ref{f:sim_refl_box_line} illustrate the performance of various causal estimators on the simulated REFL study considering Zero ATE design and Nonzero ATE design, respectively. 
Similar to findings in the simulation study in Section~\ref{s:sim}, 
doubly robust estimators, in general, outperform singly robust ATE estimators. 
Using SuperLearner for treatment and outcome modeling, doubly robust estimators achieve the smallest bias and MSE compared to using GLM, or GAM. 
Overall TMLE and AIPTW share similar performance. With GAM or SuperLearner, TMLE and AIPTW may be able to achieve a nominal confidence interval by using bootstrap confidence intervals. 
DSM has a relatively larger bias compared to TMLESL, TMLE, and AIPTW. DSM may achieve a nominal confidence interval with GAM.
PENCOMP suffers from severe bias and MSE with GLM or GAM. PENCOMP has improved performance with SuperLearner in terms of bias and MSE. However, PENCOMP suffers from a severe overcoverage and a large type I error in nearly all settings.

\section{Practical Recommendations and Discussion}
\label{s:discuss}

We have reviewed multiple doubly robust estimators and conducted simulations across a broad range of data scenarios. We vary causal inference test settings by adjusting a variety of knobs in the simulations, which include nonlinearity of treatment and outcome surfaces, degree of overlap between treatment distributions as well as treatment effect heterogeneity. 
We make use of a powerful machine learning technique SuperLearner to help improve ATE estimation. 
Also, various doubly robust estimators are applied to a real-life application of fibromyalgia as an example. 
In particular, we find that incorporating machine learning with doubly robust estimators such as the TMLE gives the best overall performance. 
Although in general TMLE and AIPTW are both efficient and have the minimum asymptotic variance under the large-sample theory, under finite sample sizes TMLE tends to be more robust to data sparsity and near violations of positivity assumption because of its range-preserving procedure for the predicted outcome estimates. Similar finding has been shown in previous studies such as \cite{van2011targeted,porter2011relative,luque2018targeted,bahamyirou2019understanding}. 
DSM is robust to model misspecification as a matching estimator, but tends to have a larger MSE compared to TMLE and AIPTW. 
The regression-based PENCOMP shows the least ideal performance among all doubly robust estimators in case of model misspecification and challenging scenarios, even when pairing with SuperLearner. 
Further research is needed to demystify the performance of PENCOMP found in our simulation studies.

Our paper helps to provide guidelines for practical use of doubly robust estimators.  Based on our extensive and realistic simulations, we recommend to estimate the ATE in the following steps: 
\begin{itemize}
    \item Perform variable selection to select outcome predictors.
    \item Model the PS and the outcomes with SuperLearner, separated by the treatment group and the control group. 
    \item Estimate the ATE by applying TMLE with the estimated propensity and outcome estimates. 
    \item Use bootstrap for variance estimation of the ATE. 
\end{itemize}

This work has multiple limitations that should be noted.  
First, throughout the paper we only consider Lasso for variable selection to illustrate the importance of the variable selection procedure. There might be better ways to remove treatment predictors such as using machine learning algorithms like random forest and neural networks.   
Soft variable selection strategies may also be used where the variable selection is conducted without requiring any modeling on the outcome, and thus provides robustness against misspecification \citep{tang2021variable}. 
Secondly, our work focuses only on doubly robust methods, which in part due to their advantages in robustness over traditional methods.  However, these methods still require correct specification of at least one of the models.  
Recent research proposed the use of model averaging across many methods to improve the robustness of comparative analyses \citep{zagar2022evaluating}.  
Future work should compare the operating characteristics of doubly robust approaches to model averaging, or perhaps simply incorporating multiple doubly robust methods within the model averaging framework. 
In addition, an extension of evaluating the use of doubly robust estimators on survival data could be explored in the future.

In summary, this work has provided best practice guidance on the use of doubly robust methods for comparative analysis based on real world data.  The use of machine learning for variable selection and model development, along with estimation of treatment effects using TMLE, are found to help improved operating characteristics of doubly robust methods.

\bibliographystyle{agsm}

\bibliography{Bibliography-MM-MC}
\end{document}